# Multilayer technique developed for fabricating Nb-based single-electron devices

Alexey B. Pavolotsky[1,2], Thomas Weimann[1], Hansjörg Scherer[1], Vladimir A. Krupenin[2], Jürgen Niemeyer[1], and Alexander B. Zorin[1]

[1] Physikalisch-Technische Bundesanstalt, Bundesallee 100, Braunschweig 38116, Germany
[2] Cryoelectronics Laboratory, Moscow State University, Vorobjevi Gori, 119899 Moscow, Russia

A reliable process has been developed for the fabrication of all-Nb single-electron circuits, based on spin-on glass planarization. The process steps are the *in situ* growth of Nb/AlO$_x$/Nb sandwich, definition of the patterns of junctions, base electrodes and wiring by use of reactive ion etching and the planarization of a spin-on glass insulation between base electrode and wiring. A single electron transistor made of 0.3×0.3 μm$^2$ area junctions clearly shows the *e*-periodic Coulomb blockade modulation by a voltage applied to a gate.

Up to now, Nb tunnel junctions have remained the most popular elements of superconductive electronics. Since the invention of the Nb/AlO$_x$/Nb trilayer by Gurvich *et al.*[1], the fabrication of complex circuits of micrometer-scale junctions has been well established. In the past years, there was a strong motivation to develop a technique for the reliable fabrication of Nb Josephson junctions of smaller sizes, characterized by smaller self-capacitances, which are advantageous for application in both Josephson devices [2,3] and single electron circuits. Moreover, due to rather large values of the superconductor energy gap of Nb ($\Delta \approx$ 1.4 meV) and, hence, larger Josephson-junction coupling, the small-capacitance Nb junctions are very promising for the single electron tunneling (SET) applications in which the interplay between the Coulomb charging energy $E_C$ and the Josephson coupling energy $E_J$ is realized [4].

A number of attempts were made to fabricate submicron-scale Nb/AlO$_x$/Nb tunnel junctions. These approaches can be divided into two groups: first, the techniques in which the tunnel barrier was formed *ex situ*[5,6], and, secondly, those in which it was grown *in situ*[7-9].

The main disadvantage of the *ex situ* techniques is the difficulty of reviving a fresh Nb interface after exposing it to the room air, which results in either bad or irreproducible junction quality. The *in situ* methods are of two different kinds. One approach is based on the shadow evaporation technique[7] which is the standard method for the fabrication of submicron Al tunnel junctions. In the case of Nb, this technique is strongly restricted, limiting the film thickness to roughly 50-70 nm per layer and prohibiting sputter cleaning of the substrate prior the film evaporation. This makes the possibility of growing the high quality Nb electrodes and, thus, the junctions doubtful.

Another option is to grow *in situ* the Nb/AlO$_x$/Nb sandwich as is done for micrometer-scale devices. Then, the junction patterns should be defined by reactive ion etching (RIE) through an SiO$_2$ mask[8] or Nb regions implanted by Ga, that plays the role of an etching mask[9]. In both cases vias in the insulation between base electrode and wiring above junction pillars are opened by the planarization technique: either by chemical mechanical polishing (CMP)[8] or by a resist partial etching back[9].

In this paper we present the novel process developed by us for fabricating the circuits with small Nb junctions. The process also uses *in situ* growth of the Nb/AlO$_x$/Nb sandwich and planarization of the interlayer insulation. There are three main features as against the preceding works[8,9]:
(1) Interlayer insulation is made of spin-on glass (SOG) which is an insulating and planarizing layer in one. This allows the complicated process of partial etching back of resist and the expensive procedure of CMP to be avoided.
(2) Ge/Cr mask, which is etch resistant but easy to remove, is used for the definition of the junctions.
(3) The Al-AlO$_x$ barrier layer is wet-etched, which prevents the formation of fences at the edges of the base electrode.

The details of the fabrication process are shown in Fig. 1. In the very first step, the trilayer of Nb/Al-AlOx/Nb was grown on the thermally oxidized silicon substrate. The layer thicknesses were chosen to be 100 nm for the bottom Nb layer, 200 nm for the top one and 10 nm for the originally sputtered Al interlayer which was thermally oxidized to form an AlO$_x$ tunnel barrier. In order to achieve sufficiently high normal resistances of the junctions which are suitable for observing SET effects[10], i.e. $R_N \gg R_Q = h/e^2 \approx$ 26 kΩ, an oxygen pressure of 300 mbar and period of 6 hours at 20°C were chosen. This regime gave the barrier a specific conductance of 0.155 mS/μm$^2$.

Subsequently the spacer layer of 40 nm Ge was thermally evaporated above the sandwich. The pattern for the junctions was defined of the top niobium layer of the sandwich. The 30 nm thick Cr mask for this etching step (Fig. 1a) was patterned by lift-off of the PMMA/Copolymer mask. Then Ge spacer and the top Nb layer of the sandwich were etched through the mask in

CF$_4$ plasma at a pressure of 10 Pa, 100 W for a 15 cm diameter cathode. The barrier layer served as a natural stop for etching. The end point was also detected by laser interferometry.

The mask for the base electrode was made of negative tone e-beam sensitive AZ PN 114 resist. After exposition and development, the barrier layer was removed by wet etching in buffered hydrofluoric acid for a few seconds (Fig. 1b). After that, the bottom niobium layer of the sandwich was etched by means of RIE (Fig. 1c) in CF$_4$ plasma, at 10 Pa, 100 W through the same mask. The completely patterned base electrode with the defined junction areas after resist mask stripping is shown in Fig. 1d.

Then the base electrode with defined junction areas was planarized by SOG Accuglass®314, made by Allied Signal Inc., which was first spun at 3000 rpm, and produced the film of about 300 nm thick. The procedure of converting this (easily solvable) polymer film into a hard (SiO$_2$-like) and stable state included soft baking on a hot plate at 90°C for 3 min and subsequent e-beam curing with a dose of 3000 μC/cm$^2$.

Opening of the vias to the junction pillars, with the base electrode and its edges remaining insulated (SOG etching back), was achieved by isotropic RIE in the reactive gas mixture of SF$_6$ (10 sccm) and O$_2$ (10 sccm) with plasma parameters of 30 W, 10 Pa (Fig. 1e). The thickness of the removed SOG was controlled by observing the light reflectivity yield of the SOG film. (Since the reflectivity yield of a transparent film, like SOG, depends periodically on the layer thickness, it makes possible to detect precisely the thickness of SOG *in situ*, directly during the etching process.)

Finally, the Cr mask and Ge spacer were stripped in *Cr Etch 18* solution, prepared by Shipley (Fig. 1f). At last, niobium layer of 100 nm thickness was deposited. The wiring pattern is defined by RIE in CF$_4$ plasma, at 10Pa, 100W through the mask of AZ PN 114 resist (Fig. 1g).

An SEM image of the final structure (Fig. 1h) is shown in Fig. 2. The structure presents the chain of junctions (islands) of different size. The pair of the smallest junctions and a gate which is capacitively coupled ($C_g$) to the small island in between, form an SET transistor.

The current-voltage curves of both the chains and the transistor recorded at $T = 4.2$ K and 30 mK (see the transistor $I$-$V$ curve in Fig. 3) show the very low subgap leakage current and sharp gap feature with the proximity knee, typical for Nb junctions with low critical current density (see, e.g., Ref. 11). The gap voltage was found to be about 2.7 mV per junction. A critical current in this (smallest-junction) sample was not detected, that we attribute to its high impedance and, hence, to the effect of fluctuations.[12]

From the voltage offset of the normal-state high-voltage asymptote ($V_{off} \approx 25$ μV) we estimated[10] the total capacitance of the transistor island to be $C_\Sigma = e/V_{off} \approx 6$ fF, leading to the charging energy $E_C/k_B = e^2/2k_B C_\Sigma \approx 150$ mK. Assuming identity of the junctions and a small contribution of the self-capacitance of the island, we evaluate the junction capacitance to be $C_j = C_\Sigma/2 \approx 3$ fF. On the assumption of the specific capacitance of about 4 μF/cm$^2$ for a barrier grown under similar conditions[13], one can evaluate the effective junction area of 0.075 μm$^2$ (compare with the nominal area of 0.09 μm$^2$) As long as the charging energy was found to be rather small, we tested the SET properties at various conditions to find the most prominent Coulomb blockade modulation. Figure 4 demonstrates the effect of the gate voltage ($V_g$) modulation at fixed transport current. The magnetic field of 1 T corresponds to the non-complete suppression of superconductivity in Nb films. It can be seen that in a rather wide range of current the curves exhibit the characteristic periodic behavior with $\Delta V_g = e/C_g \approx 41$ mV, so that $C_g \approx 4$ aF. Further reduction of the junction areas should drastically improve the signal-to-noise ratio in measurements.

In conclusion, we have developed a reliable process for fabricating of high quality Nb/AlO$_x$/Nb tunnel junctions with areas of down to 0.3×0.3 μm$^2$. Even these junctions clearly demonstrate the typical superconductive and normal tunnel properties, combined with the SET behavior of the transistor made of these junctions. We believe that the principles underlying the process elaboration allow the junction size to be decreased down to the 0.1 μm scale.

This work was supported in part by the EU (MEL ARI Research Project CHARGE and SMT Research Project SETamp), the Russian Scientific Program "Physics of Solid State Nanostructures", the Russian Fund for Fundamental Research and the German BMBF.

**Figures caption**

Fig.1. Process flow diagram for the fabrication of submicron junctions using Ge/Cr mask and e-beam cured spin-on glass planarization.

Fig. 2. Micrograph of the chain of Nb/Al-AlO$_x$/Nb junctions about 1×1 µm$^2$ in size (4 leftmost junctions), 0.5×0.5 µm$^2$ (6 rightmost junctions) and 0.3×0.3 µm$^2$ (two junctions in between). The narrow strip (seen in the upper part) plays the role of a gate.

Fig. 3. (a) The quasiparticle *I-V* characteristics of the Nb transistor in superconducting (solid line) and normal (dashed line) states, caused by applying magnetic field.

Fig.4. Gate modulation curves for the Nb transistor, recorded at several values of the transport current.

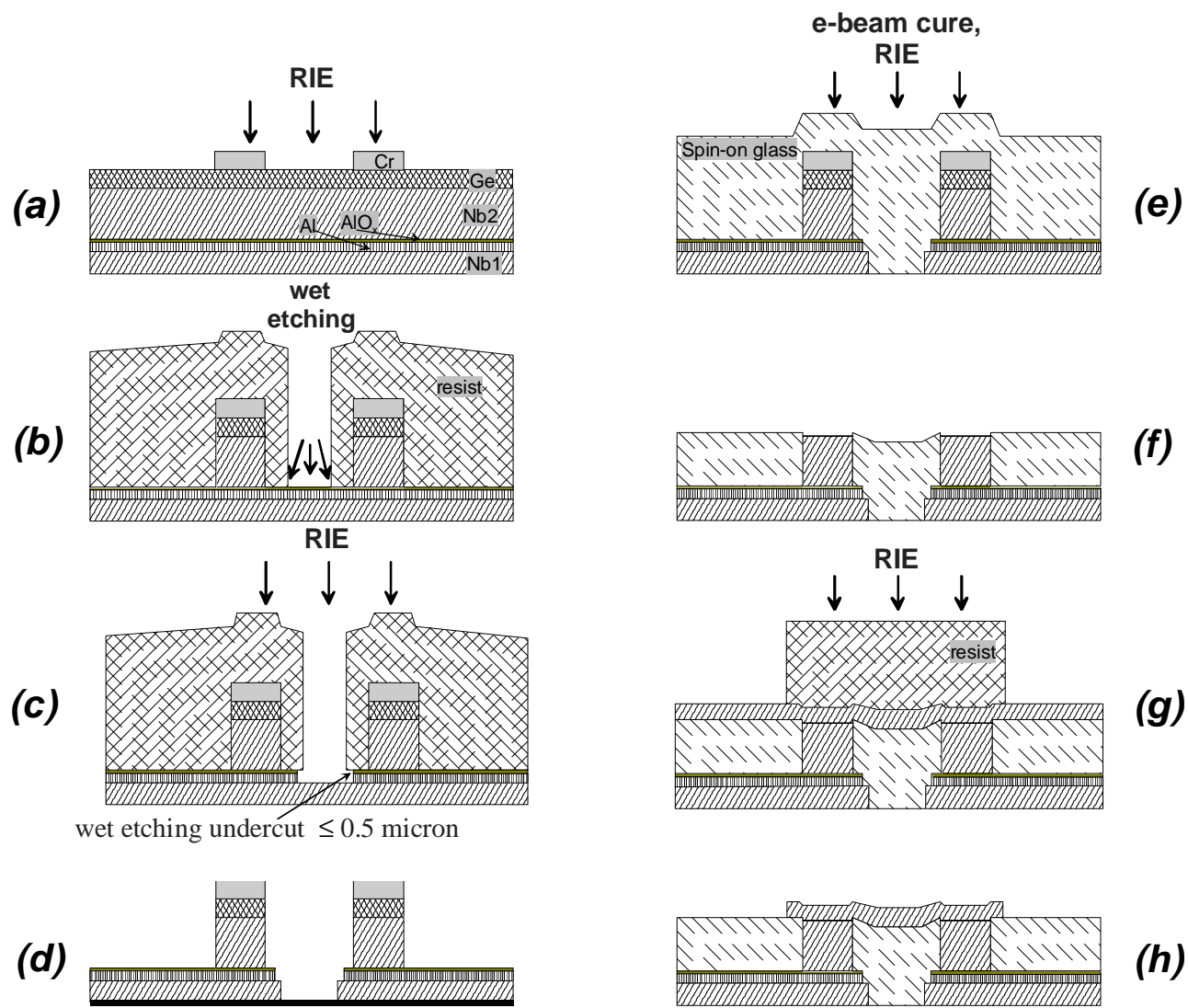

Fig. 1

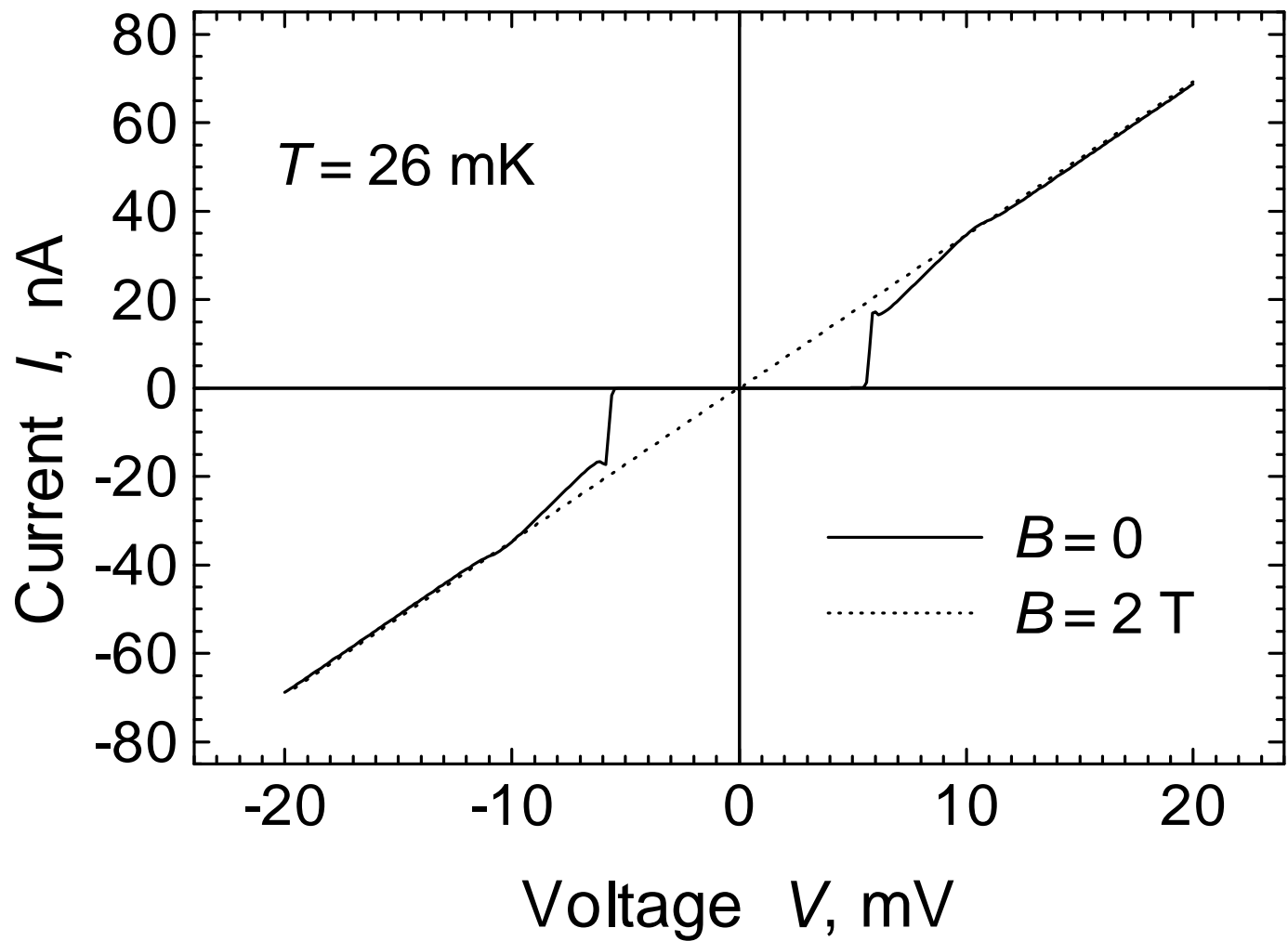

Fig.3

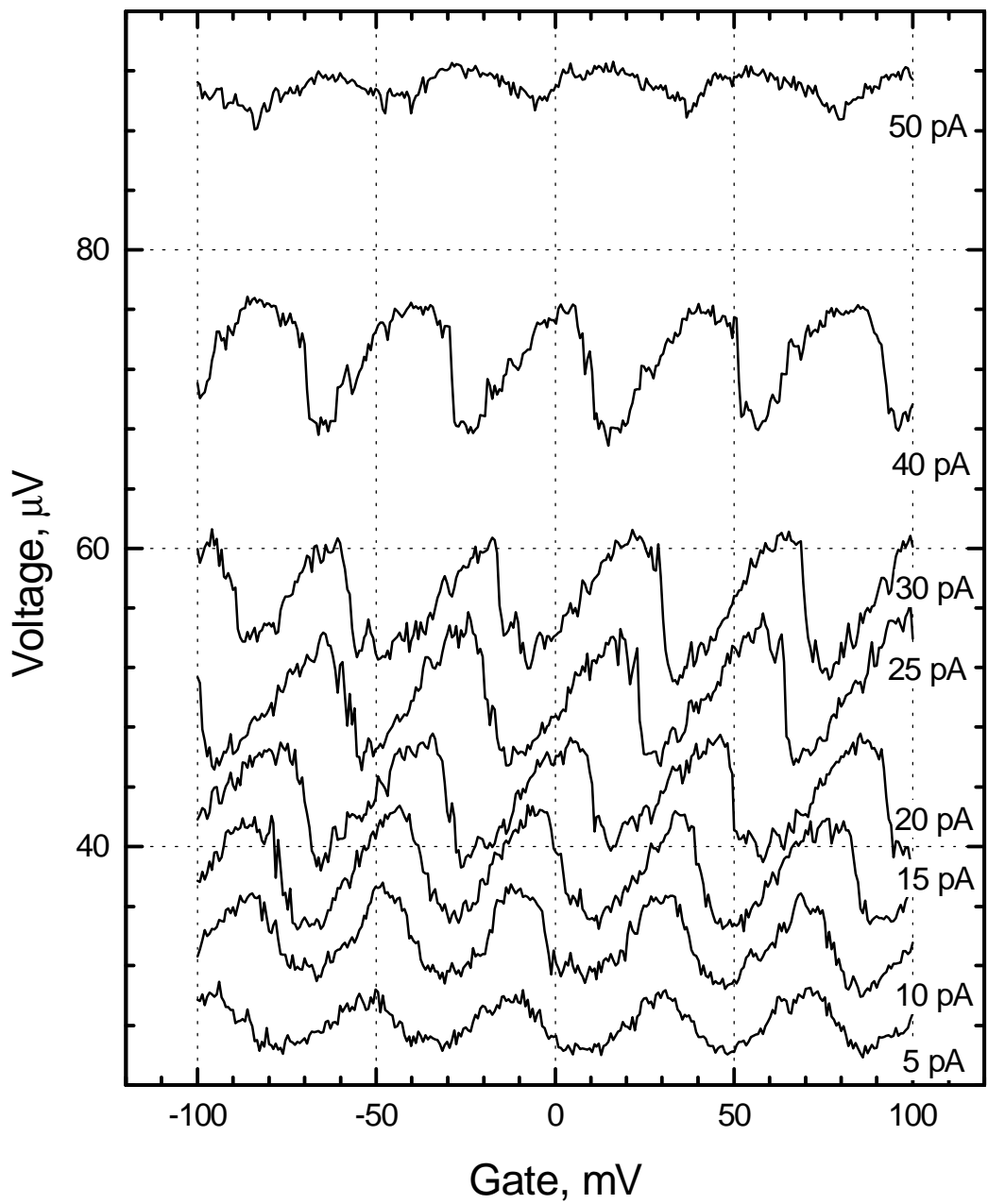

Fig. 4